\documentclass[aps,prb,reprint,showpacs,superscriptaddress,twocolumn]{revtex4-1}
\usepackage{amsmath}
\usepackage{amssymb}
\usepackage{graphicx}
\usepackage[hypertex]{hyperref}
\usepackage{amsfonts}
\usepackage{amssymb}
\usepackage{natbib}
\usepackage[usenames]{color}
\usepackage{color}
\usepackage[T1]{fontenc}
\usepackage[english]{babel}

\hypersetup{
    colorlinks=true,linkcolor=blue,citecolor=blue,
    filecolor=blue,urlcolor=blue,breaklinks=true
}
\RequirePackage{color}

\begin{document}

\title{Landau quantization, Aharonov-Bohm effect and two-dimensional pseudoharmonic quantum dot around a screw dislocation}
\author{Cleverson Filgueiras}
\email{cleverson.filgueiras@dfi.ufla.br}
\affiliation{
  Departamento de F\'{i}sica,
  Universidade Federal de Lavras, Caixa Postal 3037,
  37200-000, Lavras-MG, Brazil
}
\author{Mois\'{e}s Rojas}
\email{moises.leyva@dfi.ufla.br}
\affiliation{
  Departamento de F\'{i}sica,
  Universidade Federal de Lavras, Caixa Postal 3037,
  37200-000, Lavras-MG, Brazil
}
\author{Gilson Aciole}
\affiliation{Unidade Acadêmica de Física, Universidade Federal de Campina Grande, POB
10071, 58109-970, Campina Grande-PB, Brazil}
\author{Edilberto O. Silva}
\email{edilberto.silva@ufma.br}
\affiliation{
  Departamento de F\'{i}sica,
  Universidade Federal do Maranh\~{a}o,
  65085-580, S\~{a}o Lu\'{i}s-MA, Brazil
}
\date{\today }

\begin{abstract}
In this paper, we investigate the influence of a screw dislocation on the
energy levels and the wavefunctions of an electron confined in a
two-dimensional pseudoharmonic quantum dot under the influence of an
external magnetic field inside a dot and Aharonov-Bohm field inside a
pseudodot. The exact solutions for energy eigenvalues and wavefunctions are
computed as functions of applied uniform magnetic field strength, Aharonov-Bohm flux, magnetic quantum number and the parameter characterizing the screw
dislocation, the Burgers vector. We investigate the modifications due to
the screw dislocation on the light interband absorption coefficient and
absorption threshold frequency. Two scenarios are possible, depending on if singular effects either manifest or not. We found that as the Burgers vector
increases, the curves of frequency are pushed up towards of the growth of
it. One interesting aspect which we have observed is that the Aharonov-Bohm flux can be tuned in order to cancel the screw effect of the model.
\end{abstract}

\pacs{73.43.Cd,73.43.Qt}
\maketitle

\section{Introduction}

\label{intro}

The study of quantum dynamics for particles in constant magnetic \cite{Book.1981.Landau} and Aharonov-Bohm (AB) flux fields \cite{PR.1959.115.485}, which are perpendicular to the plane where the particles are confined, has
been carried out over the last years. The existence of other potentials are
also included, depending on the purpose of the investigation. For example,
in Ref. \cite{PRB.1996.53.6947} an exactly soluble model to describe quantum
dots, anti-dots, one-dimensional rings and straight two-dimensional wires in
the presence of such fields was proposed. It is an ideal tool to investigate
the AB effects and the persistent currents in quantum rings, for instance.
In \cite{JMST.2007.806.155}, the exact bound-state energy eigenvalues and
the corresponding eigenfunctions for several diatomic molecular systems in a
pseudoharmonic potential were analytically calculated for any arbitrary
angular momentum. The Dirac bound states of anharmonic oscillator \cite
{AoP.2014.341.153} and the nonrelativistic molecular models \cite
{AoP.2015.353.283} under external magnetic and AB flux fields were
investigated recently. Other examples can be found elsewhere.

On the other hand, the investigation on how a screw dislocation affects
quantum phenomena in semiconductors has received considerable attention. In
the continuum limit (low energy), such works are based on the geometric
theory of defects in semiconductors developed by Katanaev and Volovich \cite
{AoP.1992.216.1}. In this approach, the semiconductor with a screw
dislocation is described by a Riemann-Cartan manifold where the torsion is
associated to the Burgers vector. In this continuum limit, a screw
dislocation affects a quantum system like an isolated magnetic flux tube,
causing an AB interference phenomena \cite{ZPB.1978.29.101,Bueno}. The energy
spectrum of electrons around this kind of defect shows a profile similar to
that of the AB system \cite{PRB.1999.59.13491,PRL.1998.80.2257,EPL.1999.45.279,PLA.2012.376.2838,PLA.2012.376.2281,PLA.2008.372.3894}. These works describe the effect due to the geometric electron motion only. A
second ingredient plays an important role in these quantum systems. It is an
additional \textit{deformed potential} induce by a lattice distortion \cite
{AdP.1999.8.181}. It is a repulsive scalar potential(noncovariant) and shows
pronounced influences in the physical quantities in such systems. The impact
of this potential was first addressed in Ref. \cite{PRB.1999.59.13491},
where the scattering of electrons around a screw dislocation was
investigated. Recently, it was showed that a single screw dislocation has
profound influences on the electronic transport in semiconductors \cite
{SSC.2014.177.61}. Both contributions, the covariant and noncovariant terms,
were taken into account. For the electronic device industry, these defects
represents a problem since they interfere in the electronic properties of
the materials by way of scattering, due to such repulsive potential.
Therefore, research on screw dislocation and how it may influence the
dynamics of carriers is important for the improvement of electronic
technology, the discovery of new phenomena and better control of
transmission processes\cite{PRB1,PRB2}.

In this paper, we investigate how the quantum dots and antidots, with the
pseudoharmonic interaction and under the influence of external magnetic and
AB flux fields, are influenced by the presence of a screw dislocation. We obtain exact analytical expressions for the energy spectrum and wavefunctions. The modification due such topological defect in the light
absorption coefficient is examined and its influences in the threshold
frequency value of absorption coefficient are addressed. Two scenarios are possible, depending on if singular effects are taking into account or are not. It is found that
when the Burgers increases, the curves of such frequency are pushed up
towards its growth. It is also noted that the AB flux can be tuned in order
to cancel the influence of the screw dislocation in those physical
quantities.

The plan of this work is the following. In Sec. \ref{sec:II}, we derive the
Schrodinger equation for an electron around a screw dislocation in the
presence of an external magnetic field, an AB field and in the presence of a
two-dimensional pseudoharmonic potential. This case can find applications in
the context of quantum dots and anti-dots. In Sec. \ref{sec:III}, we
investigate how the screw dislocation affects such energy levels and we
investigate the impact on them due to the deformed potential. We consider
the electron confined on an interface so that we can discus our results in
the context of a (quasi) two dimensional electron gas (2DEG). In Sec. \ref
{sec:IV}, we investigate the modifications due to the screw dislocation on
the light interband absorption coefficient and absorption threshold
frequency. The conclusions remarks are outlined in Section \ref{sec:V}.

\section{The Schrodinger equation for an electron around a screw dislocation}

\label{sec:II}

Consider the model consisting of a non interacting
electron gas around an infinitely long linear screw dislocation oriented along the $z$-axis. The three-dimensional geometry of this medium is characterized by a
torsion which is identified with the surface density of the Burgers vector
in the classical theory of elasticity. In order to understand the dynamics of this system in a more consistent manner, we must take into account the existence
of a deformed potential, which is induced by elastic deformations on the 3D crystal.
The metric of the medium with this kind of defect is given (in cylindrical coordinates) by \cite{AoP.1992.216.1}
\begin{equation}
ds^{2}=\left( dz+\beta d\varphi \right) ^{2}+d\rho ^{2}+\rho ^{2}d\varphi
^{2},  \label{3dmetric}
\end{equation}
with $\left( \rho ,\varphi ,z\right) \rightarrow \left(
\rho ,\varphi +2\pi ,z\right) $ and $\beta $ is a parameter related to the
Burgers vector $b$ by $\beta =b/2\pi $. The induced metric describes a flat
medium with a singularity at the origin. The only non-zero component of the
torsion tensor is given by the two form
\begin{figure}[h]
\includegraphics[height=3.5cm]{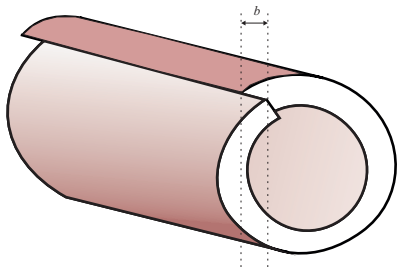}\label{f5}
\caption{{\protect\scriptsize Cylindrical portion of a 3D solid showing the
dislocation.}}\label{deslocation}
\end{figure}
\begin{equation}
T^{1}=2\pi \beta \delta ^{2}(\rho )d\rho \wedge d\varphi,  \label{curv}
\end{equation}
with $\delta ^{2}(\rho )$ being the two-dimensional delta function in the
flat space. Figure \ref{deslocation} illustrates the formation of a screw dislocation in the
bulk of a 3D crystal.

Since we consider electrons on common semiconductors, we have to introduce a
\textit{deformed potential} which describes the effects of the lattice
deformation on the electronic properties in such materials \cite
{AdP.1999.8.181}. For a screw dislocation, it is found to be
\begin{equation}
V_{d}(\rho )=\frac{\hbar ^{2}}{2ma^{2}}\frac{b^{2}}{4\pi ^{2}\rho ^{2}}\left[
2+a^{2}\left( \frac{\partial }{\partial z}\right) ^{2}\right] ,
\label{deformedV}
\end{equation}
where $a$ is the lattice constant.

The Hamiltonian for a quantum charged particle in a background $g_{ij}$
in the presence of the potential described above and in the presence of
magnetic fields is given by
\begin{equation}
H=\frac{1}{2m}\left( p_{i}-eA_{i}\right) g^{ij}\left( p_{j}-eA_{j}\right)
+V_{d}(\rho ),
\end{equation}
where $g\equiv \det {g_{ij}}$, with $i,j=\rho, \varphi, z$ and $e$ is the electric charge. For the field configuration, we consider the existence of a constant magnetic field along the $z$-direction, $\mathbf{B_{1}}=B\mathbf{\hat{z}}$, which is obtained from the potential (in the Landau gauge),
\begin{equation}
\mathbf{A}_{1}=\frac{B\rho }{2}\mathbf{\hat{\varphi}}.  \label{Al}
\end{equation}
We also consider in the model the presence of the AB potential,
\begin{equation}
\mathbf{A}_{2}=\frac{\phi _{AB}}{2\pi \rho}\mathbf{\hat{\varphi}},
\label{Aab}
\end{equation}
which provides the magnetic flux tube
\begin{equation}\label{abf}
    \mathbf{B_{2}}=\frac{\phi _{AB}}{2\pi }\frac{\delta(\rho)}{\rho}\mathbf{\hat{z}},
\end{equation}
and a scalar pseudoharmonic interaction defined by
\begin{equation}
V_{\mathrm{conf}}=V_{0}\left(\frac{\rho}{\rho_{0}}-\frac{\rho_{0}}{\rho }%
\right) ^{2},  \label{eq:}
\end{equation}
where $\rho _{0}$ and $V_{0}$ are the zero point (effective radius) and the chemical potential \cite{PB.2012.407.4198}.
As pointed out in Ref. \cite{SSC.2014.177.61}, the presence of a screw
dislocation causes an effective vector potential $\mathbf{A}_{\mathrm{eff}}$
defined by
\begin{equation}
\mathbf{A}_{\mathrm{eff}}\equiv \mathbf{\hat{\varphi}}\frac{\hbar b}{2\pi
\rho }\frac{\partial}{\partial z}.
\end{equation}
The magnitude of screw dislocation, $b$, plays a similar role to $\phi_{AB}$
in the AB system, but $\mathbf{A}_{\mathrm{eff}}$ is a differential operator
instead.

Our goal is to solve the problem of an electron gas interacting with $\mathbf{A}_{1}+
\mathbf{A}_{2}+\mathbf{A}_{\mathrm{eff}}$. This model is described by the Schrödinger equation
\begin{align}
&-\frac{\hbar^{2}}{2m}\left[\frac{\partial^{2}}{\partial z^{2}}+\frac{
\partial^{2}}{\partial \rho ^{2}}+\frac{1}{\rho}\frac{\partial}{\partial
\rho}+\frac{1}{\rho^{2}}\left( \frac{\partial}{\partial \varphi}-\beta
\frac{\partial}{\partial z}+\delta \right) ^{2}\right] \psi \notag \\
&+\left[ \frac{ieB\hbar}{2m}\left( \frac{\partial}{\partial \varphi}
-\beta \frac{\partial}{\partial z}+\delta \right) +\frac{e^{2}B^{2}\rho^{2}
}{8m}\right] \psi \notag \\
&+V_{d}(\rho)\psi+V_{\mathrm{conf}}(\rho )\psi =E\psi,  \label{scho1}
\end{align}
where $\delta \equiv e\phi_{AB}/2\pi\hbar$ is the flux parameter.

In the next section, we solve the Schrodinger equation above and
discuss the impact of such deformed potential on the energy levels around
this kind of defect.

\section{Influence of a screw dislocation on the energy levels}

\label{sec:III}

In this section, we start by investigating the influence of the screw
dislocation on the energy levels of electrons on a 3D solid taking into
account the deformed potential (\ref{deformedV}). The equation (\ref{scho1})
is solved by considering the ansatz $\psi (\rho, \varphi, z)=Ce^{il\varphi
}e^{ik_{z}z}R(\rho )$, where $l=0,\pm 1,\pm 2,\pm 3,...$, $k_{z}\in \Re $
and $C$ is a normalization constant. Equation (\ref{scho1}) leads to
\begin{align}
&-\frac{\hbar^{2}}{2m}\left[ \frac{d^{2}}{d\rho ^{2}}+\frac{1}{\rho }\frac{
d}{d\rho }-\frac{1}{\rho^{2}}\left(l-\beta k+\delta \right)^{2}-k^{2}
\right] R(\rho )  \notag \\
& +\left[ \frac{e^{2}B^{2}\rho ^{2}}{8m}-\frac{eB\hbar }{2m}\left( l-\beta
k+\delta \right) +V_{0}\left( \frac{\rho }{\rho _{0}}-\frac{\rho_{0}}{\rho}
\right) ^{2}\right] R(\rho )  \notag \\
& +\frac{\hbar ^{2}}{2ma^{2}}\frac{b^{2}}{4\pi ^{2}\rho ^{2}}\left(
2+a^{2}k^{2}\right) R(\rho )=ER(\rho ).
\end{align}
This differential equation can be rewritten as
\begin{equation}
\left[\frac{d^{2}}{d\rho^{2}}+\frac{1}{\rho}\frac{d}{d\rho}-\frac{\nu
^{2}}{\rho^{2}}-\Omega^{2}\rho ^{2}+\epsilon \right] R(\rho)=0,
\label{scho2}
\end{equation}
where
\begin{equation*}
\epsilon =\frac{2m}{\hbar^{2}}\left[E-k^{2}+\frac{eB\hbar}{2m}\left(
l-\beta k+\delta \right) +2V_{0}\right],
\end{equation*}
\begin{equation*}
\nu ^{2}=\left(l-\beta k+\delta \right)^{2}+\frac{2\beta^{2}}{a^{2}}
\left(1-\frac{k^{2}a^{2}}{2}\right)+\frac{2mV_{0}\rho_{0}^{2}}{\hbar^{2}},
\end{equation*}
and
\begin{equation*}
\Omega^{2}=\frac{e^{2}B^{2}}{4\hbar^{2}}+\frac{2mV_{0}}{\hbar^{2}\rho
_{0}^{2}}.
\end{equation*}

The general solution of the eigenvalue equation (\ref{scho2}) is given by
\cite{Book.1972.Abramowitz}
\begin{align}
R(\rho)& =C_{1}e^{-\frac{\Omega}{2}\rho ^{2}}(\Omega \rho ^{2})^{\frac{1}{2%
}+\frac{\left\vert \nu \right\vert}{2}}\mathrm{M}\left(d_{\nu
},1+\left\vert \nu \right\vert ,\Omega \rho ^{2}\right) \notag \\
& +C_{2}e^{-\frac{\Omega}{2}\rho^{2}}(\Omega \rho ^{2})^{\frac{1}{2}+\frac{%
\left\vert \nu \right\vert}{2}}\mathrm{U}\left( d_{\nu },1+\left\vert \nu
\right\vert ,\Omega \rho^{2}\right),  \label{generalsol}
\end{align}
with
\begin{equation}
d_{\nu}=\frac{1}{2}+\frac{\left\vert \nu \right\vert }{2}-\frac{\epsilon }{%
4\Omega}.  \label{arg1}
\end{equation}
The functions $\mathrm{M}$ and $\mathrm{U}$ in Eq. (\ref{generalsol}) denote
the confluent hypergeometric functions of the first and second kind,
respectively. Unlike $\mathrm{M}(a,b,z)$, which is an entire function of $z$, $U(a,b,z)$ may show a singularity at zero. If our system does not show
singularity, then we can make $C_{2}\equiv 0$. However, if the wavefunction
couples with singular potentials, we can instead, make $C_{1}\equiv 0$ \cite
{PRA.2008.77.036101}. Remember that our electron gas is in a region where
exist a topological defect, which may introduce a singularity in our
problem. We first consider the case for regular wavefunctions and after that
we discuss what changes whenever we have the irregular ones. Therefore, we
are left with
\begin{equation}
R(\rho)=C_{1}e^{-\frac{\Omega}{2}\rho^{2}}(\Omega \rho^{2})^{\frac{1}{2}+
\frac{\left\vert \nu \right\vert}{2}}\mathrm{M}\left( d_{\nu},1+\left\vert
\nu \right\vert, \Omega \rho^{2}\right).  \label{solution}
\end{equation}
A necessary condition for $R(\rho )$ to be square-integrable is $\lim_{\rho
\rightarrow \infty }R(\rho )=0$, which is fulfilled if $d_{\nu }=-n$, with $n=0,1,2,3,...$. In this way, the eigenvalues of Eq. (\ref{scho2}) are given by
\begin{equation}
E_{n}=\left( 2n+\left\vert \nu \right\vert +1\right) \frac{\Omega \hbar^{2}
}{m}-\left( l-\beta k+\delta \right) \frac{\hbar \omega _{c}}{2}+\frac{\hbar
^{2}k^{2}}{2m}-2V_{0},  \label{energy}
\end{equation}
where $\omega _{c}\equiv eB/m$ is the cyclotron frequency. Equation (\ref{energy}) are the modified Landau levels. For $\delta =0$ and $V_{0}=0$, we have
\begin{equation}
E_{n}\equiv E_{n}^{\mathrm{deff}}=\left( n+\frac{\left\vert \nu \right\vert
}{2}-\frac{l-\beta k}{2}+\frac{1}{2}\right) \hbar \omega _{c}+\frac{\hbar
^{2}k^{2}}{2m}.
\end{equation}
Notice that the deformed potential has a pronounced influence on the energy
levels. In the absence of such interaction, the energy levels are given by
\cite{EPL.1999.45.279}
\begin{equation}
E_{n}=\left( n+\frac{\left\vert l-\beta k\right\vert }{2}-\frac{l-\beta k}{2}
+\frac{1}{2}\right) \hbar \omega_{c}+\frac{\hbar ^{2}k^{2}}{2m}.
\end{equation}
For $\delta =0$, $V_{0}=0$, $\beta =0$ and ignoring the motion along the $z-$
direction, we have
\begin{equation*}
E_{n}=\left( n+\frac{1}{2}\right) \hbar \omega_{c}\;,
\end{equation*}
which are the usual Landau levels for electrons on a flat sample.

At this point, we consider the electrons on an flat interface, with
thickness $d$, around a screw dislocation. They are confined by an infinite
square well potential in the $z$-direction ($0\leq z\leq d$). This way, we
have $k_{z}=\ell \pi /d$ , where $\ell =1,2,3,...$ We will consider just the
first transverse mode $\ell =1$ filled. Then, the parameter $\nu ^{2}$ in
Eq. (\ref{scho2}) can be put in the following way,
\begin{equation}
\nu^{2}=\left(l-\frac{b}{2d}+\delta \right)^{2}+\frac{b^{2}}{2a^{2}\pi
^{2}}-\frac{1}{4}\frac{b^{2}}{d^{2}}+\frac{2mV_{0}\rho_{0}^{2}}{\hbar^{2}}.
\label{nunu}
\end{equation}
In the case where $\delta \equiv 0$ and $V_{0}\equiv 0$, we have
\begin{equation}
\nu^{2}=\left(l-\frac{b}{2d}\right)^{2}+\frac{b^{2}}{2a^{2}\pi^{2}}
-\left(\frac{b}{2d}\right)^{2}.  \label{nununu}
\end{equation}
As pointed out above, the presence of a deformed potential has a pronounced
influence on the energy levels, since in its absence, we have just
\begin{equation}
\nu =\left(l-\frac{a}{2d}\right) .
\end{equation}

In the absence of both the magnetic fields, we find
\begin{equation}
E_{n}=\hbar \sqrt{\frac{8V_{0}}{m\rho_{0}^{2}}}\left(n+\frac{1}{2}+\frac{%
\left\vert \mu \right\vert}{2}\right) +\frac{\hbar^{2}k^{2}}{2m}-2V_{0},
\label{energy2}
\end{equation}
where
\begin{equation*}
\mu^{2}=\left(l-\frac{b}{2d}\right)^{2}+\frac{b^{2}}{2a^{2}\pi^{2}}-
\frac{b^{2}}{4d^{2}}+\frac{2mV_{0}\rho_{0}^{2}}{\hbar^{2}}.
\end{equation*}
In the absence of the deformed potential, we have
\begin{equation}
\mu^{2}=\left(l-\frac{b}{2d}\right)^{2}+\frac{2mV_{0}\rho_{0}^{2}}{\hbar^{2}}.
\end{equation}

We now turn our attention to the case considering the irregular solution
which is achieved by considering $C_{1}\equiv 0$ in Eq. (\ref{generalsol}),
that is
\begin{equation}
R(\rho )=C_{1}e^{-\frac{\Omega }{2}\rho ^{2}}\left( \Omega \rho ^{2}\right)
^{\frac{1}{2}+\frac{\left\vert \nu \right\vert }{2}}\mathrm{U}\left(
-n,1+\left\vert \nu \right\vert ,\Omega \rho ^{2}\right) .  \label{solution2}
\end{equation}
As showed in Ref. \cite{AoP.2010.325.2529}, this solution diverges as $\rho
\rightarrow 0$ but is square integrable when
\begin{equation}
-1<\left\vert \nu \right\vert <1\;.  \label{constraint}
\end{equation}
The eigenvalue of (\ref{solution2}) are the same given by Eq. (\ref{energy})
but with the constraint (\ref{constraint}) above, which can be achieved only
if $l=0$. In Eq. (\ref{energy}), any value of $l$ is allowed.

\section{Influence of the screw dislocation on the interband light
absorption coefficient and in the threshold frequency value of absorption}

\label{sec:IV}

In this section, we calculate the direct interband light absorption
coefficient $K(\omega )$ and the threshold frequency of absorption in a
quantum pseudodot under the influence of external magnetic field, AB flux
field and the screw dislocation. The light absorption coefficient can be
expressed as\cite%
{PE.2004.22.860,PE.2006.31.83,PB.2005.363.262,SSS.2010.12.1253}
\begin{align}
K(\omega )& =N  \notag \\
& \times \sum_{n,l,\nu }\sum_{n^{\prime },l^{\prime },\nu ^{\prime
}}\left\vert \int \psi _{n,m,\nu }^{e}\left( \rho ,\phi \right) \psi
_{n^{\prime },l^{\prime },\nu ^{\prime }}^{h}\left( \rho ,\phi \right) \rho
d\rho d\phi \right\vert ^{2}  \notag \\
& \times \delta \left( \Delta -E_{n,m,\nu }^{e}-E_{n^{\prime },l^{\prime
},\nu ^{\prime }}^{h}\right) ,  \label{absor}
\end{align}%
where $\Delta \equiv \hbar \varpi-\varepsilon _{g}$, $\varepsilon _{g}$ is the width of forbidden energy gap, $\varpi$ is the frequency of
incident light, $N$ is a quantity proportional to the square of dipole
moment matrix element modulus, $\psi ^{e\left( h\right) }$ is the wave
function of the electron(hole) and $E^{e(h)}$ is the corresponding energy of
the electron (hole). Considering the solution (\ref{solution}), the Eq.(\ref
{absor}) becomes \cite{PB.2012.407.4198}
\begin{align}
K(\omega )& =N\sum_{n,l,\nu }\sum_{n^{\prime },l^{\prime },\nu ^{\prime }}
\frac{\Omega ^{\left\vert \nu \right\vert +\left\vert \nu ^{\prime
}\right\vert +2}(n+\left\vert \nu \right\vert )!(n^{\prime }+\left\vert \nu
^{\prime }\right\vert )!}{\pi ^{2}n!n^{\prime }!\left( \left\vert \nu
\right\vert !\right) ^{2}(\left\vert \nu ^{\prime }\right\vert !)^{2}}
\notag \\
& \times \Bigg\vert\int_{0}^{2}\pi e^{i(l+l^{\prime })}\int_{0}^{\infty
}\rho d\rho e^{-\frac{1}{2}(\Omega +\Omega ^{^{\prime }})\rho ^{2}}\rho
^{\left\vert \nu \right\vert +\left\vert \nu ^{\prime }\right\vert }  \notag
\\
& \times \,\mathrm{M}\left( -n,1+\left\vert \nu \right\vert ,\Omega \rho
^{2}\right) \mathrm{M}\left( -n^{\prime },1+\left\vert \nu ^{\prime
}\right\vert ,\Omega ^{\prime }\rho ^{2}\right) \Bigg\vert^{2}  \notag \\
& \times \,\delta \left( \Delta -E_{n,m,\nu }^{e}-E_{n^{\prime },l^{\prime
},\nu ^{\prime }}^{h}\right) .  \label{absor2}
\end{align}%
Following Ref. \cite{PB.2012.407.4198}, the light absorption coefficient is given by
\begin{equation}
K\left( \omega \right) =N\sum_{n,l,\nu}\sum_{n^{\prime},l^{\prime},\nu
^{\prime }}P_{nn^{\prime}}^{\nu}Q_{nn^{\prime }}^{\nu}\delta \left(
\Delta -E_{n,m,\nu}^{e}-E_{n^{\prime},l^{\prime},\nu^{\prime
}}^{h}\right) ,  \label{absor3}
\end{equation}%
where
\begin{align*}
P_{nn^{^{\prime}}}^{\nu}& =\frac{1}{\left(\left\vert \nu \right\vert
!\right)^{4}}(\Omega \Omega^{\prime})^{\left\vert \nu \right\vert
+1}\left(\frac{\Omega+\Omega^{\prime}}{\Omega-\Omega^{\prime}}\right)
^{2\left(n+n^{\prime}\right)} \\
& \times \frac{\left( n+\left\vert \nu \right\vert \right) !\left(n^{\prime
}+\left\vert \nu \right\vert \right) !}{n!n^{\prime}!},
\end{align*}
and
\begin{equation*}
Q_{nn^{\prime }}^{\nu }=\left[ A_{\left\vert \nu \right\vert ,\Omega }
\mathrm{M}\left( n,n^{\prime },\left\vert \nu \right\vert +1;-\frac{4\Omega
\Omega ^{\prime }}{\left( \Omega -\Omega ^{\prime }\right) ^{2}}\right)
\right] ^{2},
\end{equation*}
where
\begin{equation*}
A_{\left\vert \nu \right\vert ,\Omega }=\left( \left\vert \nu \right\vert
\right) !\left( \frac{2}{\Omega +\omega ^{\prime }}\right) ^{2\left\vert \nu
\right\vert +1}.
\end{equation*}
For the case considering the irregular solution above, we must consider
the expressions above but changing the function $\mathrm{M}$(the confluent hypergeometric functions of the first kind) by $\mathrm{U}$
(the confluent hypergeometric functions of the second kind). For this last case, only $l=0$ is allowed.

From Eqs. (\ref{energy}) and (\ref{absor2}), the threshold frequency of
absorption will be given by
\begin{align}
\hbar \varpi& =\left( 2n+\left\vert \nu \right\vert +1\right) \left(
\frac{\Omega _{h}}{m_{h}}+\frac{\Omega _{e}}{m_{e}}\right) \hbar^{2}  \notag
\\
& -\frac{1}{2}\left( \frac{1}{m_{h}}+\frac{1}{m_{e}}\right) \left(l-\beta
k+\delta \right) \hbar eB  \notag \\
& +\frac{1}{2}\left( \frac{1}{m_{h}}+\frac{1}{m_{e}}\right) \hbar
^{2}k^{2}+\varepsilon _{g}-4V_{0},  \label{eha}
\end{align}
where $m_{e}$($m_{h}$) are the electron effective mass(hole effective mass)
and
\begin{equation*}
\Omega _{e}=\sqrt{\frac{e^{2}B^{2}}{4\hbar ^{2}}+\frac{2m_{e}V_{0}}{\hbar
^{2}\rho _{0}^{2}}},\;\;\Omega _{h}=\sqrt{\frac{e^{2}B^{2}}{4\hbar ^{2}}+
\frac{2m_{h}V_{0}}{\hbar ^{2}\rho _{0}^{2}}}.
\end{equation*}
By ignoring the motion along $z$-direction and in the absence of both the
defect($\beta \equiv 0$) and the AB flux field($\delta \equiv 0$), we
recover the threshold frequency of absorption found in Ref. \cite
{SSS.2010.12.1253}. Further, taking $(n,l)=(0,0)$ in the presence of the
fields (\ref{Al}) and (\ref{Aab}), we find
\begin{align}
\hbar \varpi_{00}& =\left( \left\vert \nu _{l=0}\right\vert +1\right)
\left( \frac{\Omega _{h}}{m_{h}}+\frac{\Omega _{e}}{m_{e}}\right) \hbar ^{2}
\notag \\
& +\frac{1}{2}\left( \frac{1}{m_{h}}-\frac{1}{m_{e}}\right) \left( \delta
-\beta k\right) \hbar eB  \notag \\
& +\varepsilon _{g}+\frac{1}{2}\left( \frac{1}{m_{h}}+\frac{1}{m_{e}}\right)
\hbar ^{2}k^{2}-4V_{0}.  \label{ehb}
\end{align}
In the absence of screw dislocation, we recover the expressions of Ref. \cite
{PB.2012.407.4198}, that is
\begin{align}
\hbar \varpi_{00}^{b\equiv 0}& =\left( 1+\sqrt{\delta ^{2}+\frac{
2m_{h}V_{0}\rho _{0}^{2}}{\hbar ^{2}}}\right) \frac{\Omega _{h}}{m_{h}}\hbar
^{2}  \notag \\
& +\left( 1+\sqrt{\delta ^{2}+\frac{2m_{e}V_{0}\rho _{0}^{2}}{\hbar ^{2}}}
\right) \frac{\Omega _{e}}{m_{e}}\hbar ^{2} \notag\\
& +\varepsilon _{g}+\frac{1}{2}\left( \frac{1}{m_{h}}-\frac{1}{m_{e}}\right)
\hbar eB\delta -4V_{0}.  \label{ehc}
\end{align}
Due to the presence of the defect, we now have the following expression for
the the threshold frequency,
\begin{align}
\hbar \varpi_{00}^{\mathrm{screw}}& =\left( 1+\sqrt{\delta ^{2}-\frac{
\delta b}{d}+\frac{b^{2}}{2a^{2}\pi ^{2}}+\frac{2m_{e}V_{0}\rho _{0}^{2}}{
\hbar ^{2}}}\right) \frac{\Omega _{e}}{m_{e}}\hbar ^{2}  \notag \\
& +\left( 1+\sqrt{\delta ^{2}-\frac{\delta b}{d}+\frac{b^{2}}{2a^{2}\pi ^{2}}
+\frac{2m_{h}V_{0}\rho _{0}^{2}}{\hbar ^{2}}}\right) \frac{\Omega _{h}}{m_{h}
}\hbar ^{2}  \notag \\
& +\varepsilon _{g}+\frac{1}{2}\left( \frac{1}{m_{h}}-\frac{1}{m_{e}}\right)
\left( \delta -\frac{b}{2d}\right) \hbar eB-4V_{0},  \label{eh2}
\end{align}
if the deformed potential is taken into account. When $\delta \equiv 0$, Eq.
(\ref{eh2}) become\begin{align}
\hbar \varpi_{00}^{\mathrm{screw}}& =\left( 1+\sqrt{\frac{b^{2}}{2a^{2}\pi ^{2}}+\frac{2m_{e}V_{0}\rho _{0}^{2}}{\hbar ^{2}}}\right) \frac{\Omega _{e}}{m_{e}}\hbar ^{2}  \notag \\
& +\left( 1+\sqrt{\frac{b^{2}}{2a^{2}\pi ^{2}}+\frac{2m_{h}V_{0}\rho _{0}^{2}}{\hbar ^{2}}}\right) \frac{\Omega _{h}}{m_{h}}\hbar ^{2}  \notag \\
& +\varepsilon _{g}-\frac{1}{4d}\left( \frac{1}{m_{h}}-\frac{1}{m_{e}}\right) \hbar eBb-4V_{0}.
\end{align}
Let us now investigate the influence of the screw dislocation on the
threshold value of absorption for transition $000\rightarrow 000$, comparing
the two cases, one with and the other without the noncovariant deformed
potential. The argument of Dirac delta function allows one to define such
threshold value of absorption as
\begin{equation*}
\frac{\hbar \varpi_{00}}{\varepsilon _{g}}=1+\frac{E_{0}^{e}+E_{0}^{h}
}{\varepsilon _{g}}.
\end{equation*}
Let us now set the following parameters:
\begin{equation*}
\eta =\frac{\rho _{0}\varepsilon _{g}}{\hbar }\sqrt{\frac{m_{e}}{V_{0}}}
,\;\;\eta ^{\prime }=\frac{\rho _{0}\varepsilon _{g}}{\hbar }\sqrt{\frac{m_{h}
}{V_{0}}},
\end{equation*}
\begin{equation*}
\kappa =\frac{e\hbar B}{m_{e}\varepsilon _{g}},\;\kappa ^{\prime }=\frac{
e\hbar B}{m_{h}\varepsilon _{g}}.
\end{equation*}
For a quantum dot, we take $V_{0}\rho _{0}^{2}\rightarrow 0$. In this case,
we obtain
\begin{equation}
\frac{E_{0}^{e}}{\varepsilon _{g}}=\frac{1}{2}(1+\xi )\sqrt{\kappa ^{2}+
\frac{8}{\eta ^{2}}}-\frac{\kappa }{2}\left( \delta -\frac{b}{2d}\right)-
\frac{2V_{0}}{\varepsilon _{g}},
\end{equation}
\begin{equation}
\frac{E_{0}^{h}}{\varepsilon _{g}}=\frac{1}{2}(1+\xi )\sqrt{\kappa ^{\prime
2}+\frac{8}{\eta ^{\prime 2}}}+\frac{\kappa ^{\prime }}{2}\left( \delta -
\frac{b}{2d}\right) -\frac{2V_{0}}{\varepsilon _{g}},
\end{equation}
where
\begin{figure}[b]
\includegraphics[scale=0.29]{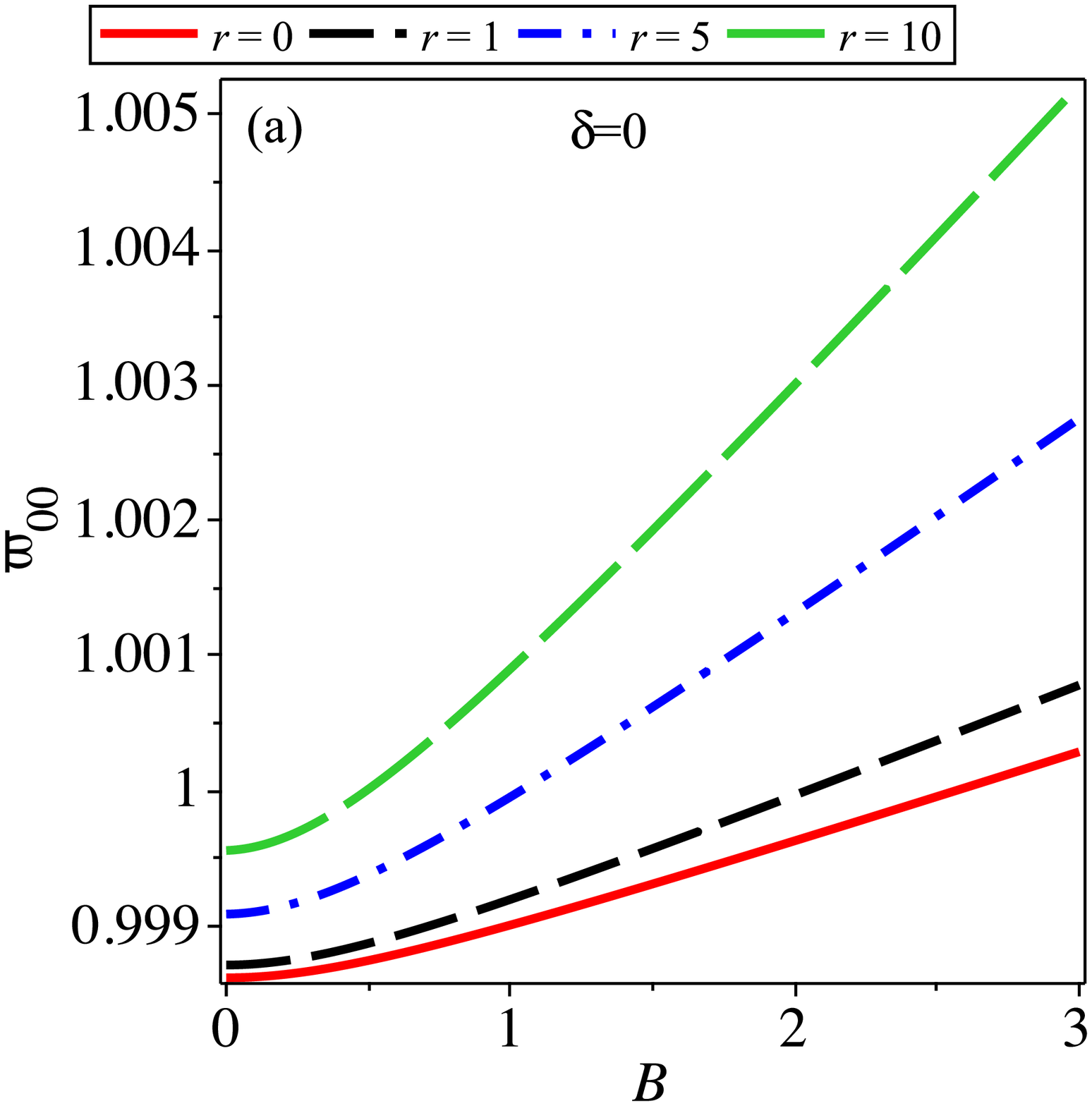}\\
\includegraphics[scale=0.29]{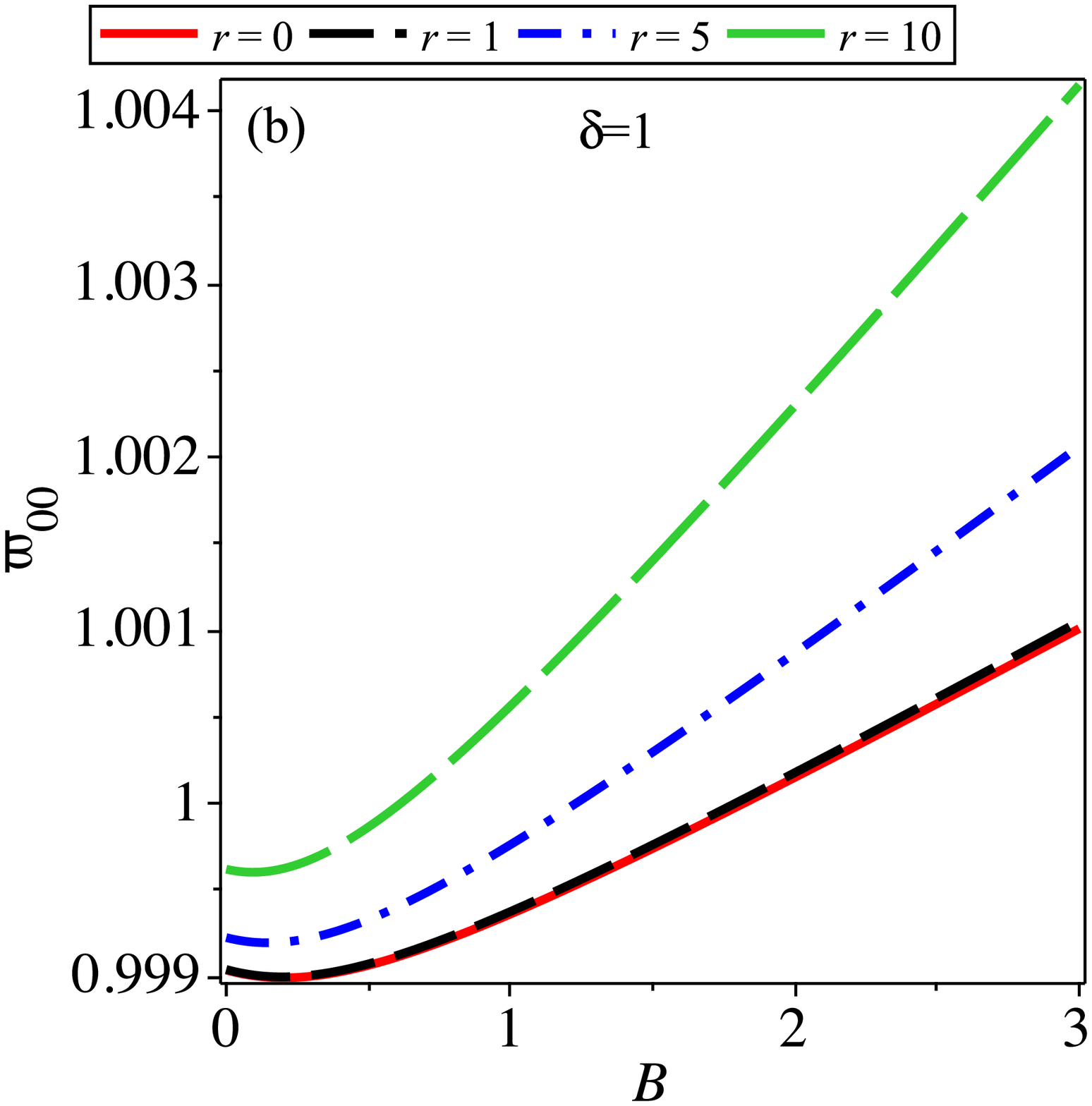}
\caption{The variations of threshold frequency of absorption $\protect\varpi
_{00}$ (in unit of $\protect\epsilon_{g}$) as function of $B$ for several
values of the $r$ parameter. In (a) we display $\protect\varpi_{00}$ for
absence of a flux field $\protect\delta=0$ and in (b) we display $\protect
\varpi_{00}$ for $\protect\delta=1.0$.}
\label{fig:w2a}
\end{figure}
\begin{equation*}
\xi =\sqrt{\delta ^{2}-\frac{\delta b}{d}+\frac{b^{2}}{2a^{2}\pi ^{2}}}.
\end{equation*}
On the other hand, for a quantum anti-dot, we take the limit $\left(
V_{0}/\rho _{0}^{2}\right) \rightarrow 0$. In this case, by maintaining the deformed potential, we have
\begin{figure}[b]
\includegraphics[scale=0.30]{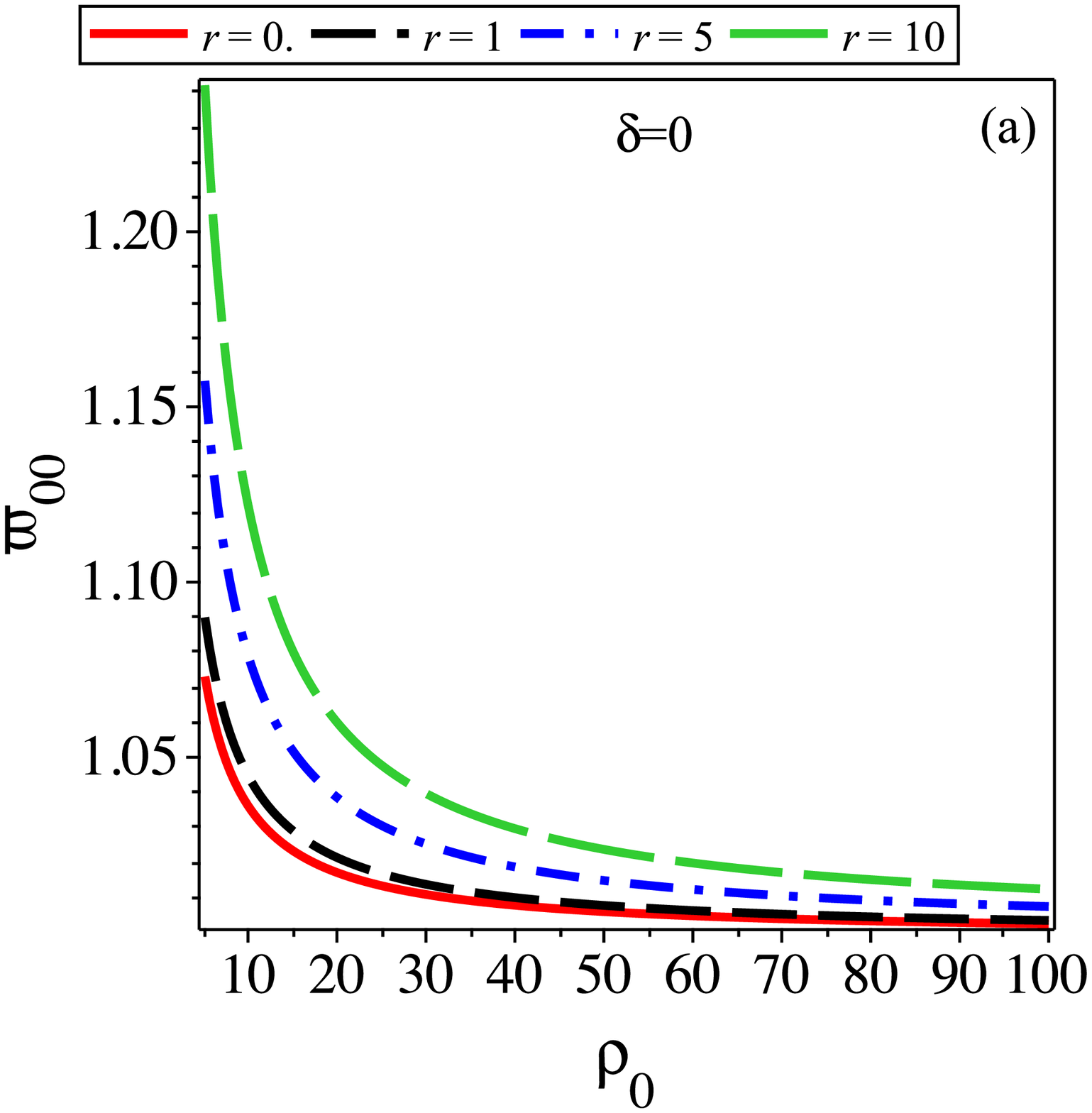}\\
\includegraphics[scale=0.30]{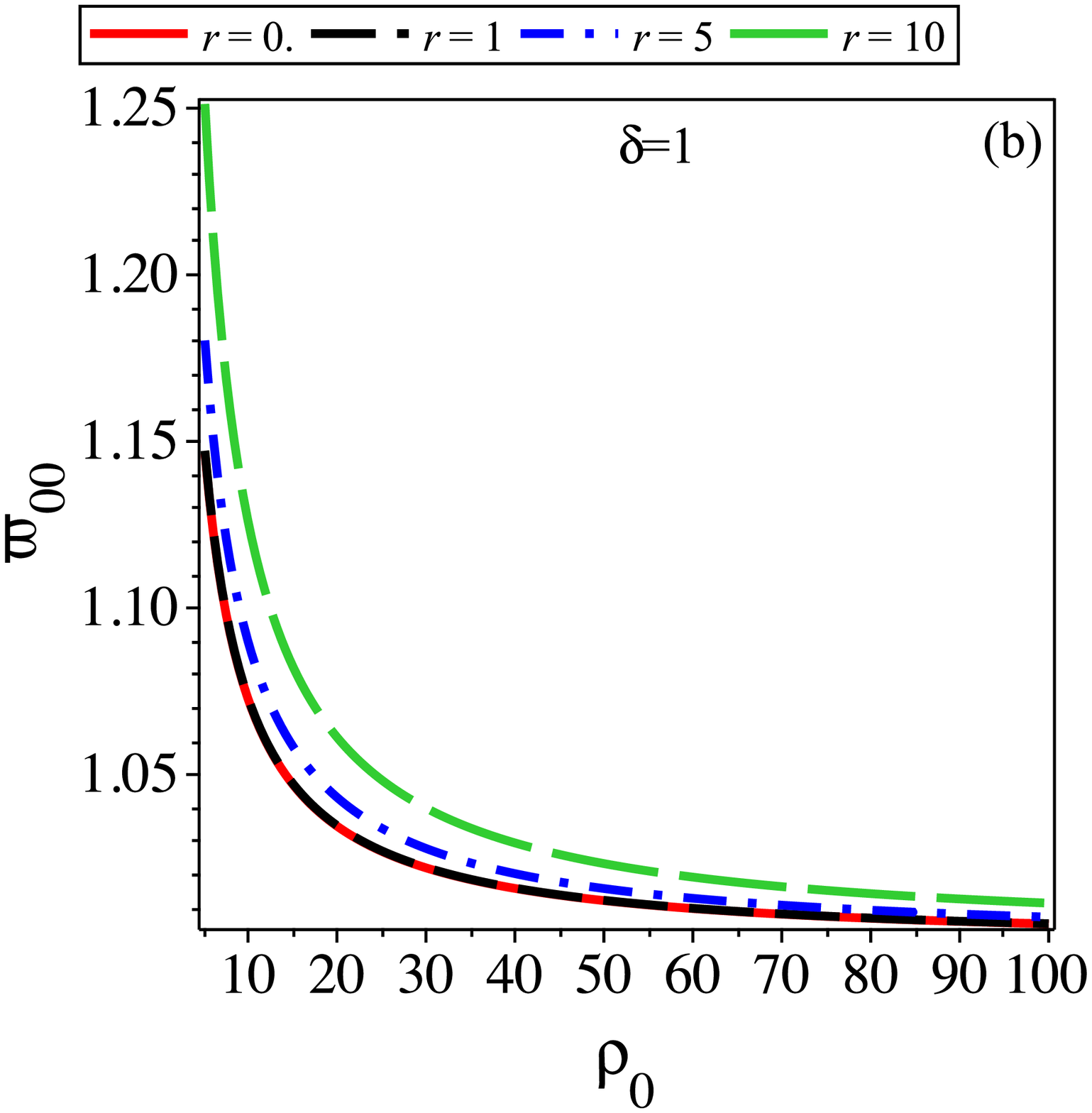}
\caption{The variations of threshold frequency of absorption $\protect\varpi
_{00}$ (in unit of $\protect\epsilon_{g}$) as function of $\protect\rho_{0}$
for several values of the $r$ parameter. In (a) we depicted $\protect\varpi
_{00}$ for $\protect\delta=0$ and in (b) $\protect\varpi_{00}$
is plotted for $\protect\delta=1.0$.}
\label{fig:w2b}
\end{figure}
\begin{figure}[b]
\includegraphics[scale=0.30]{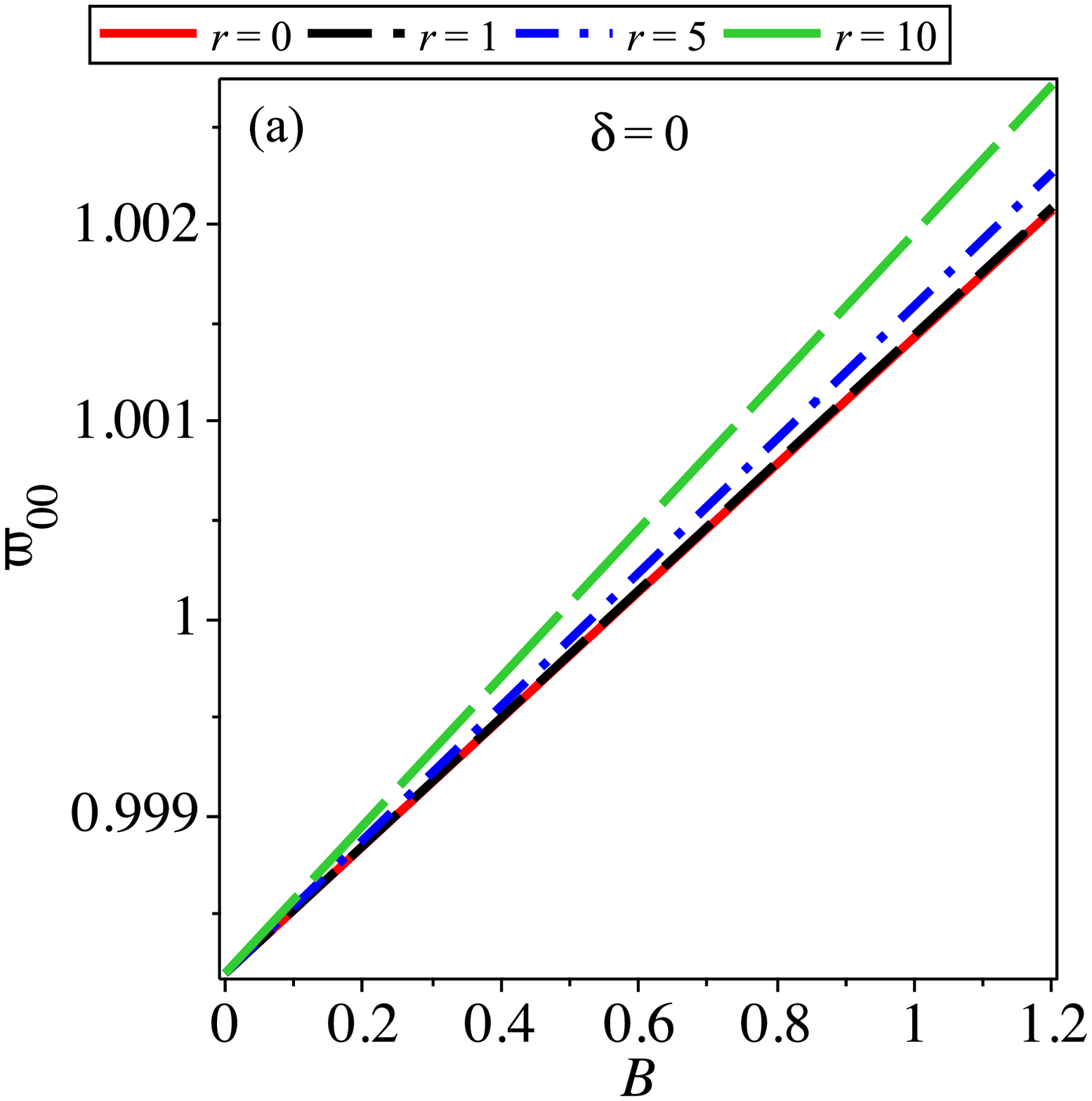}\\
\includegraphics[scale=0.30]{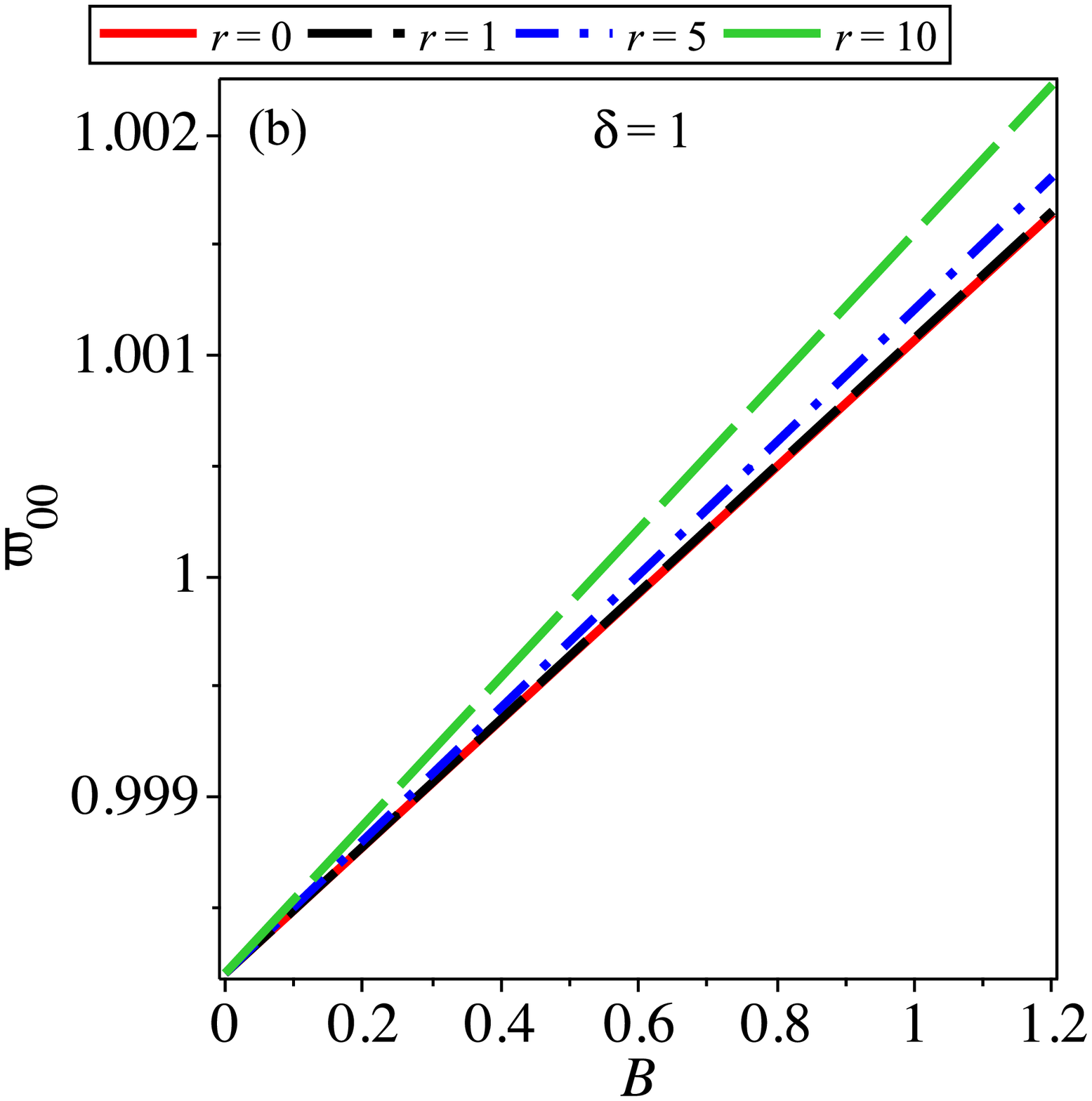}
\caption{The anti-dot frequency of absorption $\protect\varpi_{00}$ as
function of $B$ for several values of the $r$ parameter. In (a) we display $
\protect\varpi_{00}$ for absence of a flux field $\protect\delta=0$. In (b) we
display $\protect\varpi_{00}$ for $\protect\delta=1.0$}
\label{fig:w2c}
\end{figure}
\begin{equation}
\frac{E_{0}^{e}}{\varepsilon _{g}}=\frac{\kappa }{2}\left( \varsigma
+1-\delta +\frac{b}{2d}\right) -\frac{2V_{0}}{\varepsilon _{g}},\label{dot1}
\end{equation}
\begin{equation}
\frac{E_{0}^{h}}{\varepsilon _{g}}=\frac{\kappa ^{\prime }}{2}\left(
\varsigma +1+\delta -\frac{b}{2d}\right) -\frac{2V_{0}}{\varepsilon _{g}},\label{dot2}
\end{equation}
where
\begin{equation*}
\varsigma =\sqrt{\delta ^{2}-\frac{\delta b}{d}+\frac{b^{2}}{2a^{2}\pi ^{2}}+
\frac{2m_{e}V_{0}\rho _{0}^{2}}{\hbar ^{2}}}.
\end{equation*}
Now we study the effect of the AB flux field, the presence and absence of Burgers vector, quantum dot potential and quantum anti-dot potential on the threshold frequency of absorption $\varpi_{00}$ for the typical 2D structure of GaAs, with the following parameters: $V_{0}=0.68346$ meV, $a=5${\AA}, $d=15$nm, $m_{e}=0.067m_{0}$ and $m_{h}=0.34m_{0}$, where $m_{0}$ is the free electron effective mass.
In Fig. \ref{fig:w2a}, we show the variations of the threshold frequency of
absorption $\varpi_{00}$ (in units of $\varepsilon _{g}$) for a
quantum dot as a function of magnetic field (in units of Tesla) for
different values of the ratio $\lambda =b/2d$, which characterize the
influence of a screw dislocation and the parameter $r=b/a$.

In Fig. \ref{fig:w2a}(a) shows the variations of the threshold frequency of
absorption $\varpi_{00}$ at a fixed energy gap $\varepsilon _{g}$, the
chemical potential $V_{0}$ and AB flux $\delta =0$, for four
values of positive $\lambda $ and $r$. It is seen that $\varpi_{00}$ increases when the applied magnetic field increases. It is easily seen from the figure that the dependence of $\varpi_{00}$ on $B$ is nonlinear for small applied magnetic fields. On the other hand, by increasing the magnetic field the lines remain linear. It is also noted that when $r$ increases, the curves of frequency $\varpi
_{00}$ are pushed up towards of the growth $r$.

The Fig. \ref{fig:w2a}(b) illustrates the behavior of $\varpi_{00}$ for different values of the parameter $r$ and fixed AB flux $\delta =1$. From this figure,
we can see that the threshold frequency $\varpi_{00}$ displays a
minimum for weak magnetic field. On the other hand, for strong magnetic
field behavior is linear. One interesting aspect is that the magnetic flux $\delta $ contributes to the cancellation of screw effect of the model. In the figure this
cancellation occurs when $\delta =1$ and $r=1$.

In Fig. \ref{fig:w2b}, we have plots of the variations of the threshold frequency of
absorption $\varpi_{00}$ with quantum dot size (in units of $\rho _{0}$). It is seen in Fig. \ref{fig:w2b}(a) that $\varpi_{00}$ decreases when the
quantum dot size increases in absence of the AB flux $\delta =0$. In Fig
\ref{fig:w2b}(b), we can see that there is an overlap of the curves for $r=0$ and $r=1$
when $\delta =1$, this indicates to us that the AB flux field help to be
canceled the screw dislocation.

The effect of the AB flux field on the anti-dot threshold frequency of
absorption $\varpi_{00}$ are shown in Fig. \ref{fig:w2c}. In Fig. \ref{fig:w2c}(a), we plot the variation of the anti-dot threshold frequency of
absorption $\varpi_{00}$ in the absence of the AB flux field $\delta=0$ as a
function of magnetic field. We find that the dependence of $\varpi_{00}$ on $
B$ is linear. Fig \ref{fig:w2c}(b) demonstrates the dependence of the anti-dot threshold frequency on $B$ at $\delta=1.0$ with different values of $r$. The behavior of $\varpi_{00} $ is similar frequency of absorption of Fig. \ref{fig:w2c}(a), but the value of $\varpi_{00}$ for $\delta=1$ is always less that frequency of absorption $
\varpi_{00}$ when $\delta$ takes the value $0$.

\section{Concluding Remarks}

\label{sec:V}

In conclusion, we have investigated the energy levels for a 2DEG around a
screw dislocation under the pseudoharmonic interaction consisting of quantum
dot and anti-dot potentials in the presence of an uniform strong
magnetic field $B$ and AB flux.

It is known that such a defect on elastic media generates a torsion field
that acts on the particle as if an external AB flux were being applied to
it. In the usual AB effect, the charged particles in the presence of an
uniform magnetic field may be confined to a plane perpendicular to the field
lines. This is not possible in the case of torsion, which needs motion in
three-dimensional space in order to show up its effects. Because of this
fact, we have considered a quasi two-dimensional electron gas confined on a thin
interface in such a way that the effects of torsion can manifest. Due to the
existence of a screw dislocation, we have considered the effects of two
contributions: a covariant term, which comes from the geometric approach in
the continuum limit and a noncovariant repulsive scalar potential. Both
appear due to elastic deformations on a semiconductor with such kind of
topological defect. We have found that this noncovariant term changes
significantly the energy levels of electrons in this system. As we have said
above, these defects represents a problem since they interfere in the
electronic properties of the materials by way of scattering, as we can note
by analyzing the non covariant potential in Eq. (\ref{deformedV}). Therefore, investigations on how this kind of defect
influences the dynamics of carriers in common semiconductors are important
for the improvement of electronic technology. In our case, the modification
introduced by such topological defect in the light absorption coefficient is
due to an effective angular momentum induced by torsion. For the threshold
frequency value of absorption, we have found that, as the Burgers vector
increases, the curves of such frequency are pushed up towards its growth. Moreover,
it was noted that the singular effects can take place as well. It
is also noted that the AB flux can be tuned in order to cancel the
influence of the screw dislocation in these physical quantities.

\section*{Acknowledgments}

This work was supported by the Brazilian agencies CNPq, FAPEMA and FAPEMIG.

\bibliographystyle{apsrev4-1}

\begin{thebibliography}{27}%
\makeatletter
\providecommand \@ifxundefined [1]{%
 \@ifx{#1\undefined}
}%
\providecommand \@ifnum [1]{%
 \ifnum #1\expandafter \@firstoftwo
 \else \expandafter \@secondoftwo
 \fi
}%
\providecommand \@ifx [1]{%
 \ifx #1\expandafter \@firstoftwo
 \else \expandafter \@secondoftwo
 \fi
}%
\providecommand \natexlab [1]{#1}%
\providecommand \enquote  [1]{``#1''}%
\providecommand \bibnamefont  [1]{#1}%
\providecommand \bibfnamefont [1]{#1}%
\providecommand \citenamefont [1]{#1}%
\providecommand \href@noop [0]{\@secondoftwo}%
\providecommand \href [0]{\begingroup \@sanitize@url \@href}%
\providecommand \@href[1]{\@@startlink{#1}\@@href}%
\providecommand \@@href[1]{\endgroup#1\@@endlink}%
\providecommand \@sanitize@url [0]{\catcode `\\12\catcode `\$12\catcode
  `\&12\catcode `\#12\catcode `\^12\catcode `\_12\catcode `\%12\relax}%
\providecommand \@@startlink[1]{}%
\providecommand \@@endlink[0]{}%
\providecommand \url  [0]{\begingroup\@sanitize@url \@url }%
\providecommand \@url [1]{\endgroup\@href {#1}{\urlprefix }}%
\providecommand \urlprefix  [0]{URL }%
\providecommand \Eprint [0]{\href }%
\providecommand \doibase [0]{http://dx.doi.org/}%
\providecommand \selectlanguage [0]{\@gobble}%
\providecommand \bibinfo  [0]{\@secondoftwo}%
\providecommand \bibfield  [0]{\@secondoftwo}%
\providecommand \translation [1]{[#1]}%
\providecommand \BibitemOpen [0]{}%
\providecommand \bibitemStop [0]{}%
\providecommand \bibitemNoStop [0]{.\EOS\space}%
\providecommand \EOS [0]{\spacefactor3000\relax}%
\providecommand \BibitemShut  [1]{\csname bibitem#1\endcsname}%
\let\auto@bib@innerbib\@empty
\bibitem [{\citenamefont {Landau}\ and\ \citenamefont
  {Lifschitz}(1981)}]{Book.1981.Landau}%
  \BibitemOpen
  \bibfield  {author} {\bibinfo {author} {\bibfnamefont {L.~D.}\ \bibnamefont
  {Landau}}\ and\ \bibinfo {author} {\bibfnamefont {E.~M.}\ \bibnamefont
  {Lifschitz}},\ }\href@noop {} {\emph {\bibinfo {title} {Quantum Mechanics}}}\
  (\bibinfo  {publisher} {Pergamon},\ \bibinfo {address} {Oxford},\ \bibinfo
  {year} {1981})\BibitemShut {NoStop}%
\bibitem [{\citenamefont {Aharonov}\ and\ \citenamefont
  {Bohm}(1959)}]{PR.1959.115.485}%
  \BibitemOpen
  \bibfield  {author} {\bibinfo {author} {\bibfnamefont {Y.}~\bibnamefont
  {Aharonov}}\ and\ \bibinfo {author} {\bibfnamefont {D.}~\bibnamefont
  {Bohm}},\ }\href {\doibase 10.1103/PhysRev.115.485} {\bibfield  {journal}
  {\bibinfo  {journal} {Phys. Rev.}\ }\textbf {\bibinfo {volume} {115}},\
  \bibinfo {pages} {485} (\bibinfo {year} {1959})}\BibitemShut {NoStop}%
\bibitem [{\citenamefont {Tan}\ and\ \citenamefont
  {Inkson}(1996)}]{PRB.1996.53.6947}%
  \BibitemOpen
  \bibfield  {author} {\bibinfo {author} {\bibfnamefont {W.-C.}\ \bibnamefont
  {Tan}}\ and\ \bibinfo {author} {\bibfnamefont {J.~C.}\ \bibnamefont
  {Inkson}},\ }\href {\doibase 10.1103/PhysRevB.53.6947} {\bibfield  {journal}
  {\bibinfo  {journal} {Phys. Rev. B}\ }\textbf {\bibinfo {volume} {53}},\
  \bibinfo {pages} {6947} (\bibinfo {year} {1996})}\BibitemShut {NoStop}%
\bibitem [{\citenamefont {Ikhdair}\ and\ \citenamefont
  {Sever}(2007)}]{JMST.2007.806.155}%
  \BibitemOpen
  \bibfield  {author} {\bibinfo {author} {\bibfnamefont {S.}~\bibnamefont
  {Ikhdair}}\ and\ \bibinfo {author} {\bibfnamefont {R.}~\bibnamefont
  {Sever}},\ }\href {\doibase http://dx.doi.org/10.1016/j.theochem.2006.11.019}
  {\bibfield  {journal} {\bibinfo  {journal} {J. Mol. Structure: Theochem}\
  }\textbf {\bibinfo {volume} {806}},\ \bibinfo {pages} {155 } (\bibinfo {year}
  {2007})}\BibitemShut {NoStop}%
\bibitem [{\citenamefont {Hamzavi}\ \emph {et~al.}(2014)\citenamefont
  {Hamzavi}, \citenamefont {Ikhdair},\ and\ \citenamefont
  {Falaye}}]{AoP.2014.341.153}%
  \BibitemOpen
  \bibfield  {author} {\bibinfo {author} {\bibfnamefont {M.}~\bibnamefont
  {Hamzavi}}, \bibinfo {author} {\bibfnamefont {S.~M.}\ \bibnamefont
  {Ikhdair}}, \ and\ \bibinfo {author} {\bibfnamefont {B.~J.}\ \bibnamefont
  {Falaye}},\ }\href {\doibase http://dx.doi.org/10.1016/j.aop.2013.12.003}
  {\bibfield  {journal} {\bibinfo  {journal} {Ann. Phys.}\ }\textbf {\bibinfo
  {volume} {341}},\ \bibinfo {pages} {153 } (\bibinfo {year}
  {2014})}\BibitemShut {NoStop}%
\bibitem [{\citenamefont {Ikhdair}\ \emph {et~al.}(2015)\citenamefont
  {Ikhdair}, \citenamefont {Falaye},\ and\ \citenamefont
  {Hamzavi}}]{AoP.2015.353.283}%
  \BibitemOpen
  \bibfield  {author} {\bibinfo {author} {\bibfnamefont {S.~M.}\ \bibnamefont
  {Ikhdair}}, \bibinfo {author} {\bibfnamefont {B.~J.}\ \bibnamefont {Falaye}},
  \ and\ \bibinfo {author} {\bibfnamefont {M.}~\bibnamefont {Hamzavi}},\ }\href
  {\doibase http://dx.doi.org/10.1016/j.aop.2014.11.017} {\bibfield  {journal}
  {\bibinfo  {journal} {Ann. Phys.}\ }\textbf {\bibinfo {volume} {353}},\
  \bibinfo {pages} {282 } (\bibinfo {year} {2015})}\BibitemShut {NoStop}%
\bibitem [{\citenamefont {Katanaev}\ and\ \citenamefont
  {Volovich}(1992)}]{AoP.1992.216.1}%
  \BibitemOpen
  \bibfield  {author} {\bibinfo {author} {\bibfnamefont {M.}~\bibnamefont
  {Katanaev}}\ and\ \bibinfo {author} {\bibfnamefont {I.}~\bibnamefont
  {Volovich}},\ }\href {\doibase 10.1016/0003-4916(52)90040-7} {\bibfield
  {journal} {\bibinfo  {journal} {Ann. Phys. (NY)}\ }\textbf {\bibinfo {volume}
  {216}},\ \bibinfo {pages} {1} (\bibinfo {year} {1992})}\BibitemShut {NoStop}%
\bibitem [{\citenamefont {Kawamura}(1978)}]{ZPB.1978.29.101}%
  \BibitemOpen
  \bibfield  {author} {\bibinfo {author} {\bibfnamefont {K.}~\bibnamefont
  {Kawamura}},\ }\href {\doibase 10.1007/BF01313193} {\bibfield  {journal}
  {\bibinfo  {journal} {Zeitschrift f{\"u}r Physik B Condensed Matter}\
  }\textbf {\bibinfo {volume} {29}},\ \bibinfo {pages} {101} (\bibinfo {year}
  {1978})}\BibitemShut {NoStop}%
\bibitem [{\citenamefont {Bueno}\ \emph {et~al.}(2016)\citenamefont {Bueno},
  \citenamefont {Furtado},\ and\ \citenamefont {Bakke}}]{Bueno}%
  \BibitemOpen
  \bibfield  {author} {\bibinfo {author} {\bibfnamefont {M.}~\bibnamefont
  {Bueno}}, \bibinfo {author} {\bibfnamefont {C.}~\bibnamefont {Furtado}}, \
  and\ \bibinfo {author} {\bibfnamefont {K.}~\bibnamefont {Bakke}},\ }\href
  {\doibase http://dx.doi.org/10.1016/j.physb.2016.05.026} {\bibfield
  {journal} {\bibinfo  {journal} {Physica B: Condensed Matter}\ }\textbf
  {\bibinfo {volume} {496}},\ \bibinfo {pages} {45 } (\bibinfo {year}
  {2016})}\BibitemShut {NoStop}%
\bibitem [{\citenamefont {Bausch}\ \emph
  {et~al.}(1999{\natexlab{a}})\citenamefont {Bausch}, \citenamefont {Schmitz},\
  and\ \citenamefont {Turski}}]{PRB.1999.59.13491}%
  \BibitemOpen
  \bibfield  {author} {\bibinfo {author} {\bibfnamefont {R.}~\bibnamefont
  {Bausch}}, \bibinfo {author} {\bibfnamefont {R.}~\bibnamefont {Schmitz}}, \
  and\ \bibinfo {author} {\bibfnamefont {L.~A.}\ \bibnamefont {Turski}},\
  }\href {\doibase 10.1103/PhysRevB.59.13491} {\bibfield  {journal} {\bibinfo
  {journal} {Phys. Rev. B}\ }\textbf {\bibinfo {volume} {59}},\ \bibinfo
  {pages} {13491} (\bibinfo {year} {1999}{\natexlab{a}})}\BibitemShut {NoStop}%
\bibitem [{\citenamefont {Bausch}\ \emph {et~al.}(1998)\citenamefont {Bausch},
  \citenamefont {Schmitz},\ and\ \citenamefont {Turski}}]{PRL.1998.80.2257}%
  \BibitemOpen
  \bibfield  {author} {\bibinfo {author} {\bibfnamefont {R.}~\bibnamefont
  {Bausch}}, \bibinfo {author} {\bibfnamefont {R.}~\bibnamefont {Schmitz}}, \
  and\ \bibinfo {author} {\bibfnamefont {L.~A.}\ \bibnamefont {Turski}},\
  }\href {\doibase 10.1103/PhysRevLett.80.2257} {\bibfield  {journal} {\bibinfo
   {journal} {Phys. Rev. Lett.}\ }\textbf {\bibinfo {volume} {80}},\ \bibinfo
  {pages} {2257} (\bibinfo {year} {1998})}\BibitemShut {NoStop}%
\bibitem [{\citenamefont {Furtado}\ and\ \citenamefont
  {Moraes}(1999)}]{EPL.1999.45.279}%
  \BibitemOpen
  \bibfield  {author} {\bibinfo {author} {\bibfnamefont {C.}~\bibnamefont
  {Furtado}}\ and\ \bibinfo {author} {\bibfnamefont {F.}~\bibnamefont
  {Moraes}},\ }\href {\doibase 10.1209/epl/i1999-00159-8} {\bibfield  {journal}
  {\bibinfo  {journal} {Europhys. Lett.}\ }\textbf {\bibinfo {volume} {45}},\
  \bibinfo {pages} {279} (\bibinfo {year} {1999})}\BibitemShut {NoStop}%
\bibitem [{\citenamefont {Bakke}\ and\ \citenamefont
  {Moraes}(2012)}]{PLA.2012.376.2838}%
  \BibitemOpen
  \bibfield  {author} {\bibinfo {author} {\bibfnamefont {K.}~\bibnamefont
  {Bakke}}\ and\ \bibinfo {author} {\bibfnamefont {F.}~\bibnamefont {Moraes}},\
  }\href {\doibase http://dx.doi.org/10.1016/j.physleta.2012.09.006} {\bibfield
   {journal} {\bibinfo  {journal} {Phys. Lett. A}\ }\textbf {\bibinfo {volume}
  {376}},\ \bibinfo {pages} {2838 } (\bibinfo {year} {2012})}\BibitemShut
  {NoStop}%
\bibitem [{\citenamefont {Lorenci}\ and\ \citenamefont
  {Jr.}(2012)}]{PLA.2012.376.2281}%
  \BibitemOpen
  \bibfield  {author} {\bibinfo {author} {\bibfnamefont {V.~A.~D.}\
  \bibnamefont {Lorenci}}\ and\ \bibinfo {author} {\bibfnamefont {E.~S.~M.}\
  \bibnamefont {Jr.}},\ }\href {\doibase
  http://dx.doi.org/10.1016/j.physleta.2012.05.055} {\bibfield  {journal}
  {\bibinfo  {journal} {Phys. Lett. A}\ }\textbf {\bibinfo {volume} {376}},\
  \bibinfo {pages} {2281 } (\bibinfo {year} {2012})}\BibitemShut {NoStop}%
\bibitem [{\citenamefont {Netto}\ \emph {et~al.}(2008)\citenamefont {Netto},
  \citenamefont {Chesman},\ and\ \citenamefont {Furtado}}]{PLA.2008.372.3894}%
  \BibitemOpen
  \bibfield  {author} {\bibinfo {author} {\bibfnamefont {A.~S.}\ \bibnamefont
  {Netto}}, \bibinfo {author} {\bibfnamefont {C.}~\bibnamefont {Chesman}}, \
  and\ \bibinfo {author} {\bibfnamefont {C.}~\bibnamefont {Furtado}},\ }\href
  {\doibase http://dx.doi.org/10.1016/j.physleta.2008.02.060} {\bibfield
  {journal} {\bibinfo  {journal} {Phys. Lett. A}\ }\textbf {\bibinfo {volume}
  {372}},\ \bibinfo {pages} {3894 } (\bibinfo {year} {2008})}\BibitemShut
  {NoStop}%
\bibitem [{\citenamefont {Bausch}\ \emph
  {et~al.}(1999{\natexlab{b}})\citenamefont {Bausch}, \citenamefont {Schmitz},\
  and\ \citenamefont {Turski}}]{AdP.1999.8.181}%
  \BibitemOpen
  \bibfield  {author} {\bibinfo {author} {\bibfnamefont {R.}~\bibnamefont
  {Bausch}}, \bibinfo {author} {\bibfnamefont {R.}~\bibnamefont {Schmitz}}, \
  and\ \bibinfo {author} {\bibfnamefont {u.~A.}\ \bibnamefont {Turski}},\
  }\href {\doibase
  10.1002/(SICI)1521-3889(199903)8:3<181::AID-ANDP181>3.0.CO;2-H} {\bibfield
  {journal} {\bibinfo  {journal} {Ann. Phys. (Leipzig)}\ }\textbf {\bibinfo
  {volume} {8}},\ \bibinfo {pages} {181} (\bibinfo {year}
  {1999}{\natexlab{b}})}\BibitemShut {NoStop}%
\bibitem [{\citenamefont {Taira}\ and\ \citenamefont
  {Shima}(2014)}]{SSC.2014.177.61}%
  \BibitemOpen
  \bibfield  {author} {\bibinfo {author} {\bibfnamefont {H.}~\bibnamefont
  {Taira}}\ and\ \bibinfo {author} {\bibfnamefont {H.}~\bibnamefont {Shima}},\
  }\href {\doibase http://dx.doi.org/10.1016/j.ssc.2013.10.002} {\bibfield
  {journal} {\bibinfo  {journal} {Solid State Communications}\ }\textbf
  {\bibinfo {volume} {177}},\ \bibinfo {pages} {61 } (\bibinfo {year}
  {2014})}\BibitemShut {NoStop}%
\bibitem [{\citenamefont {Slager}\ \emph {et~al.}(2014)\citenamefont {Slager},
  \citenamefont {Mesaros}, \citenamefont {Juri\ifmmode \check{c}\else
  \v{c}\fi{}i\ifmmode~\acute{c}\else \'{c}\fi{}},\ and\ \citenamefont
  {Zaanen}}]{PRB1}%
  \BibitemOpen
  \bibfield  {author} {\bibinfo {author} {\bibfnamefont {R.-J.}\ \bibnamefont
  {Slager}}, \bibinfo {author} {\bibfnamefont {A.}~\bibnamefont {Mesaros}},
  \bibinfo {author} {\bibfnamefont {V.}~\bibnamefont {Juri\ifmmode
  \check{c}\else \v{c}\fi{}i\ifmmode~\acute{c}\else \'{c}\fi{}}}, \ and\
  \bibinfo {author} {\bibfnamefont {J.}~\bibnamefont {Zaanen}},\ }\href
  {\doibase 10.1103/PhysRevB.90.241403} {\bibfield  {journal} {\bibinfo
  {journal} {Phys. Rev. B}\ }\textbf {\bibinfo {volume} {90}},\ \bibinfo
  {pages} {241403} (\bibinfo {year} {2014})}\BibitemShut {NoStop}%
\bibitem [{\citenamefont {Slager}\ \emph {et~al.}(2016)\citenamefont {Slager},
  \citenamefont {Juri\ifmmode \check{c}\else \v{c}\fi{}i\ifmmode~\acute{c}\else
  \'{c}\fi{}}, \citenamefont {Lahtinen},\ and\ \citenamefont {Zaanen}}]{PRB2}%
  \BibitemOpen
  \bibfield  {author} {\bibinfo {author} {\bibfnamefont {R.-J.}\ \bibnamefont
  {Slager}}, \bibinfo {author} {\bibfnamefont {V.}~\bibnamefont {Juri\ifmmode
  \check{c}\else \v{c}\fi{}i\ifmmode~\acute{c}\else \'{c}\fi{}}}, \bibinfo
  {author} {\bibfnamefont {V.}~\bibnamefont {Lahtinen}}, \ and\ \bibinfo
  {author} {\bibfnamefont {J.}~\bibnamefont {Zaanen}},\ }\href {\doibase
  10.1103/PhysRevB.93.245406} {\bibfield  {journal} {\bibinfo  {journal} {Phys.
  Rev. B}\ }\textbf {\bibinfo {volume} {93}},\ \bibinfo {pages} {245406}
  (\bibinfo {year} {2016})}\BibitemShut {NoStop}%
\bibitem [{\citenamefont {Ikhdair}\ and\ \citenamefont
  {Hamzavi}(2012)}]{PB.2012.407.4198}%
  \BibitemOpen
  \bibfield  {author} {\bibinfo {author} {\bibfnamefont {S.~M.}\ \bibnamefont
  {Ikhdair}}\ and\ \bibinfo {author} {\bibfnamefont {M.}~\bibnamefont
  {Hamzavi}},\ }\href {\doibase http://dx.doi.org/10.1016/j.physb.2012.07.004}
  {\bibfield  {journal} {\bibinfo  {journal} {Physica B: Condensed Matter}\
  }\textbf {\bibinfo {volume} {407}},\ \bibinfo {pages} {4198 } (\bibinfo
  {year} {2012})}\BibitemShut {NoStop}%
\bibitem [{\citenamefont {Abramowitz}\ and\ \citenamefont
  {Stegun}(1972)}]{Book.1972.Abramowitz}%
  \BibitemOpen
  \bibinfo {editor} {\bibfnamefont {M.}~\bibnamefont {Abramowitz}}\ and\
  \bibinfo {editor} {\bibfnamefont {I.~A.}\ \bibnamefont {Stegun}},\ eds.,\
  \href@noop {} {\emph {\bibinfo {title} {Handbook of Mathematical
  Functions}}}\ (\bibinfo  {publisher} {New York: Dover Publications},\
  \bibinfo {year} {1972})\BibitemShut {NoStop}%
\bibitem [{\citenamefont {Hagen}(2008)}]{PRA.2008.77.036101}%
  \BibitemOpen
  \bibfield  {author} {\bibinfo {author} {\bibfnamefont {C.~R.}\ \bibnamefont
  {Hagen}},\ }\href {\doibase 10.1103/PhysRevA.77.036101} {\bibfield  {journal}
  {\bibinfo  {journal} {Phys. Rev. A}\ }\textbf {\bibinfo {volume} {77}},\
  \bibinfo {pages} {036101} (\bibinfo {year} {2008})}\BibitemShut {NoStop}%
\bibitem [{\citenamefont {Filgueiras}\ \emph {et~al.}(2010)\citenamefont
  {Filgueiras}, \citenamefont {Silva}, \citenamefont {Oliveira},\ and\
  \citenamefont {Moraes}}]{AoP.2010.325.2529}%
  \BibitemOpen
  \bibfield  {author} {\bibinfo {author} {\bibfnamefont {C.}~\bibnamefont
  {Filgueiras}}, \bibinfo {author} {\bibfnamefont {E.~O.}\ \bibnamefont
  {Silva}}, \bibinfo {author} {\bibfnamefont {W.}~\bibnamefont {Oliveira}}, \
  and\ \bibinfo {author} {\bibfnamefont {F.}~\bibnamefont {Moraes}},\ }\href
  {\doibase 10.1016/j.aop.2010.05.012} {\bibfield  {journal} {\bibinfo
  {journal} {Ann. Phys. (NY)}\ }\textbf {\bibinfo {volume} {325}},\ \bibinfo
  {pages} {2529} (\bibinfo {year} {2010})}\BibitemShut {NoStop}%
\bibitem [{\citenamefont {Atoyan}\ \emph {et~al.}(2004)\citenamefont {Atoyan},
  \citenamefont {Kazaryan},\ and\ \citenamefont {Sarkisyan}}]{PE.2004.22.860}%
  \BibitemOpen
  \bibfield  {author} {\bibinfo {author} {\bibfnamefont {M.}~\bibnamefont
  {Atoyan}}, \bibinfo {author} {\bibfnamefont {E.}~\bibnamefont {Kazaryan}}, \
  and\ \bibinfo {author} {\bibfnamefont {H.}~\bibnamefont {Sarkisyan}},\ }\href
  {\doibase http://dx.doi.org/10.1016/j.physe.2003.09.042} {\bibfield
  {journal} {\bibinfo  {journal} {Physica E: Low-dimensional Systems and
  Nanostructures}\ }\textbf {\bibinfo {volume} {22}},\ \bibinfo {pages} {860 }
  (\bibinfo {year} {2004})}\BibitemShut {NoStop}%
\bibitem [{\citenamefont {Atoyan}\ \emph {et~al.}(2006)\citenamefont {Atoyan},
  \citenamefont {Kazaryan},\ and\ \citenamefont {Sarkisyan}}]{PE.2006.31.83}%
  \BibitemOpen
  \bibfield  {author} {\bibinfo {author} {\bibfnamefont {M.}~\bibnamefont
  {Atoyan}}, \bibinfo {author} {\bibfnamefont {E.}~\bibnamefont {Kazaryan}}, \
  and\ \bibinfo {author} {\bibfnamefont {H.}~\bibnamefont {Sarkisyan}},\ }\href
  {\doibase http://dx.doi.org/10.1016/j.physe.2005.10.008} {\bibfield
  {journal} {\bibinfo  {journal} {Physica E: Low-dimensional Systems and
  Nanostructures}\ }\textbf {\bibinfo {volume} {31}},\ \bibinfo {pages} {83 }
  (\bibinfo {year} {2006})}\BibitemShut {NoStop}%
\bibitem [{\citenamefont {Raigoza}\ \emph {et~al.}(2005)\citenamefont
  {Raigoza}, \citenamefont {Morales},\ and\ \citenamefont
  {Duque}}]{PB.2005.363.262}%
  \BibitemOpen
  \bibfield  {author} {\bibinfo {author} {\bibfnamefont {N.}~\bibnamefont
  {Raigoza}}, \bibinfo {author} {\bibfnamefont {A.}~\bibnamefont {Morales}}, \
  and\ \bibinfo {author} {\bibfnamefont {C.}~\bibnamefont {Duque}},\ }\href
  {\doibase http://dx.doi.org/10.1016/j.physb.2005.03.031} {\bibfield
  {journal} {\bibinfo  {journal} {Physica B: Condensed Matter}\ }\textbf
  {\bibinfo {volume} {363}},\ \bibinfo {pages} {262 } (\bibinfo {year}
  {2005})}\BibitemShut {NoStop}%
\bibitem [{\citenamefont {Khordad}(2010)}]{SSS.2010.12.1253}%
  \BibitemOpen
  \bibfield  {author} {\bibinfo {author} {\bibfnamefont {R.}~\bibnamefont
  {Khordad}},\ }\href {\doibase
  http://dx.doi.org/10.1016/j.solidstatesciences.2010.03.001} {\bibfield
  {journal} {\bibinfo  {journal} {Solid State Sciences}\ }\textbf {\bibinfo
  {volume} {12}},\ \bibinfo {pages} {1253 } (\bibinfo {year}
  {2010})}\BibitemShut {NoStop}%
\end{thebibliography}

\end{document}